\documentclass[twocolumn,aps,superscriptaddress]{revtex4-1}
\pdfoutput=1

\usepackage[pdftex]{graphicx}
\usepackage{amsmath}
\usepackage{hyperref}
\usepackage{mathtools}
\usepackage{eufrak}

\DeclarePairedDelimiterX\braket[2]{\langle}{\rangle}{#1 \delimsize\vert #2}

\begin{document}

\title{Twisted 3D holograms for self-referencing interferometers in
  metrology and imaging}

\author{Martin Berz} 
\affiliation{IFE Institut f\"{u}r Forschung und Entwicklung, 81675 Munich, Trogerstr. 38, Germany, martin.berz@ife-project.com}

\author{Cordelia Berz} 
\affiliation{IFE Institut f\"{u}r Forschung und Entwicklung, 81675 Munich, Trogerstr. 38, Germany, martin.berz@ife-project.com}

\date{\today}

\begin{abstract}
The interference between radiation fields superposed appropriately
contains all available information about the source. This will be
recapitulated for coherent and incoherent fields. We will further
analyze a new kind of twisted 3D interferometer which allows us to
generate interferograms with high information content. The physical
basis for these devices is the geometric parallel transport of
electric fields along a 3D path in space. This concept enables us to
build very compact 3D interferometers.
\end{abstract}

\maketitle


\section{Introduction}
\label{introduction}

In some cases, modern optical applications need an interference
between a field and a rotated copy of this field. The rotational
shearing interferometer\cite{malacara} is an example for this. More
general applications can be found in imaging and self-referencing
interferometers (SRI)~\cite{berz}\cite{berz2}.

The following common methods are known to generate such a
superposition. They are called 'component based rotations'.
\begin{itemize}
\item two lenses in a beam expander configuration, e.g. two lenses of
  focal length f in a distance 2f, which actually flip the field by
  $180^\circ$ using a Mach-Zehnder interferometer~\cite{berz}.
\item two Dove prisms in both paths of a Mach-Zehnder
  interferometer~\cite{malacara}.
\item use of retro reflectors in a Michelson
  interferometer~\cite{malacara}.
\end{itemize}

It is the purpose of this report to describe a new geometric method
for the generation of such interferograms using parallel field
transport.

\section{Parallel transport in 2D and 3D space}
\label{trans}

Field rotations due to the parallel transport of an electric field
along the light beam do not play a vital part in most optical systems. These
systems are called 'flat'. One reason is that experimentally,
the deviation from 'flatness' can only be determined if two different field
transports are compared. This is normally an interferometric setup
which has not been fully investigated yet.

The expression 'field transport' means that the field is transported
along paths in 3D space. In a first, simple approach, the term
'parallel transport' can be understood as the 'best possible
projection' of the electric field on the vector space orthogonal to
the propagation direction of light. Such a projection is necessary to
maintain the transversal wave character if the direction of light
changes. More accurate definitions of the term 'parallel transport' can be given
in differential geometry. Yet, this is not needed here since
firstly a strict application of Maxwell's equations always leads to the
correct result, and secondly we transport the electric fields by
reflections on mirror like surfaces only. Thus, the mapping from the
incoming to the outgoing field is straightforward.

\begin{figure}
\begin{center}
\includegraphics[scale=0.32]{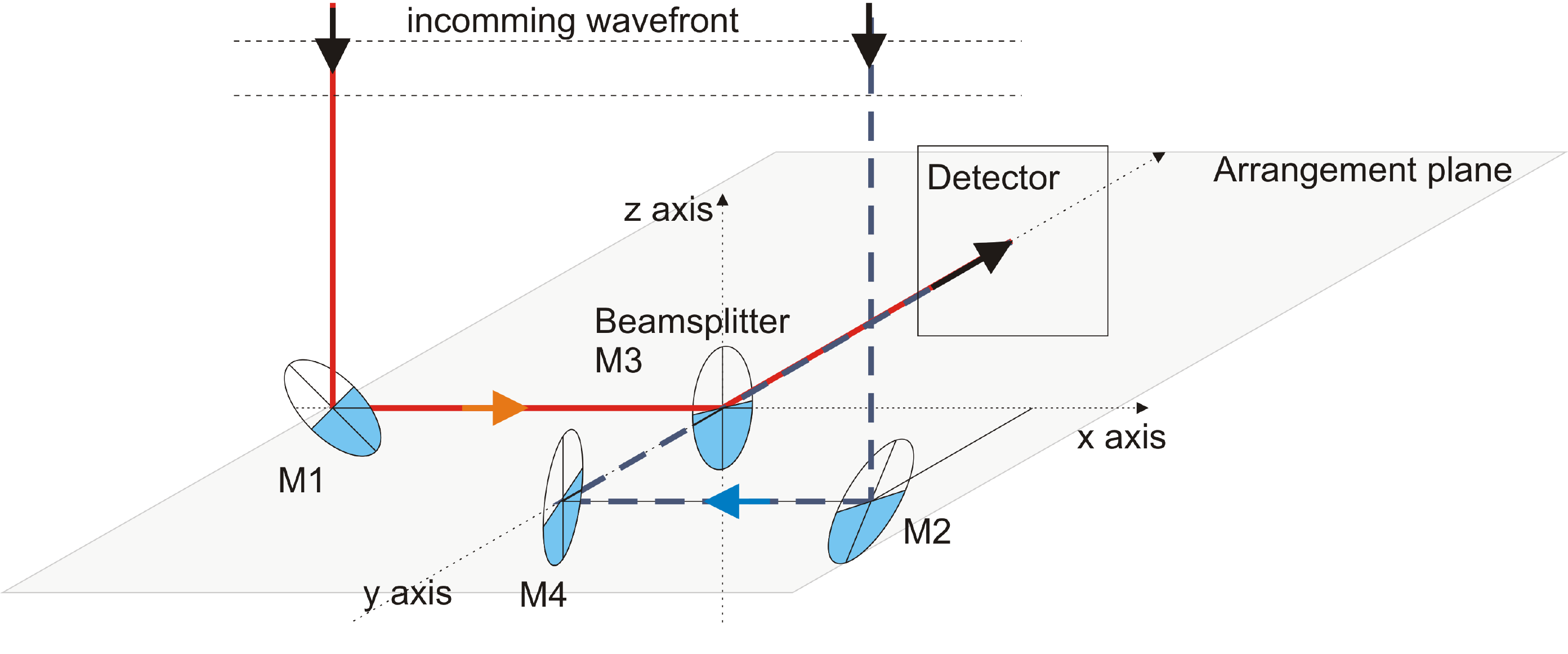}
\caption{\label{generic}One of the simplest 3D interferometer setups
  with field rotation. The interferometer uses a division of the
  wavefront, i.e. different parts of the incoming wave are fed to the
  device entries. The mirrors M1,M2,M4 and the beam splitter
  M3 are situated in the arrangement plane which is perpendicular to
  the incoming central beam. The central beam is defined by the
  property that the two beams have exactly the same direction after
  the superposing beamsplitter. The detector is perpendicular to the
  direction of the central beam. This implies that the central beam
  possesses no wave vector component in the detector plane.}
\end{center}
\end{figure}

However, the concept of parallel transport is helpful to understand the
mechanism by which the Maxwell calculus leads to mutually rotated
fields.

It immediately follows from the preceding introduction that such a
parallel transport does not give a rotation if the field paths are
purely in a 2D plane. In 2 dimensions, the rotation at the
superposition point of two paths is independent of the specific path
since rotations in 2D space commute. This behavior obviously changes
if components are inserted into the 2D path that rotate the
field. This is for instance used in a rotational shearing
interferometer where Dove prisms are inserted into the
path~\cite{malacara} (component based rotation). Outside the
components, the central light path in the two branches is still
contained in a 2D plane. Therefore, such an interferometer is called a
2D interferometer.

In contrast, a 3D interferometer is characterized by the property that
the central light paths do not lie in one and the same plane, not even
approximately.

This definition of a 2D and a 3D interferometer is independent of the fact
that all physical interferometers are actually built in a three
dimensional space.

\section{Field transport in a interferometer: 3D vs. 2D}
\label{3d_interf}

In this paper we will explore the field transport properties of
a 3D interferometer.

\begin{figure}
\begin{center}
\includegraphics[scale=0.32]{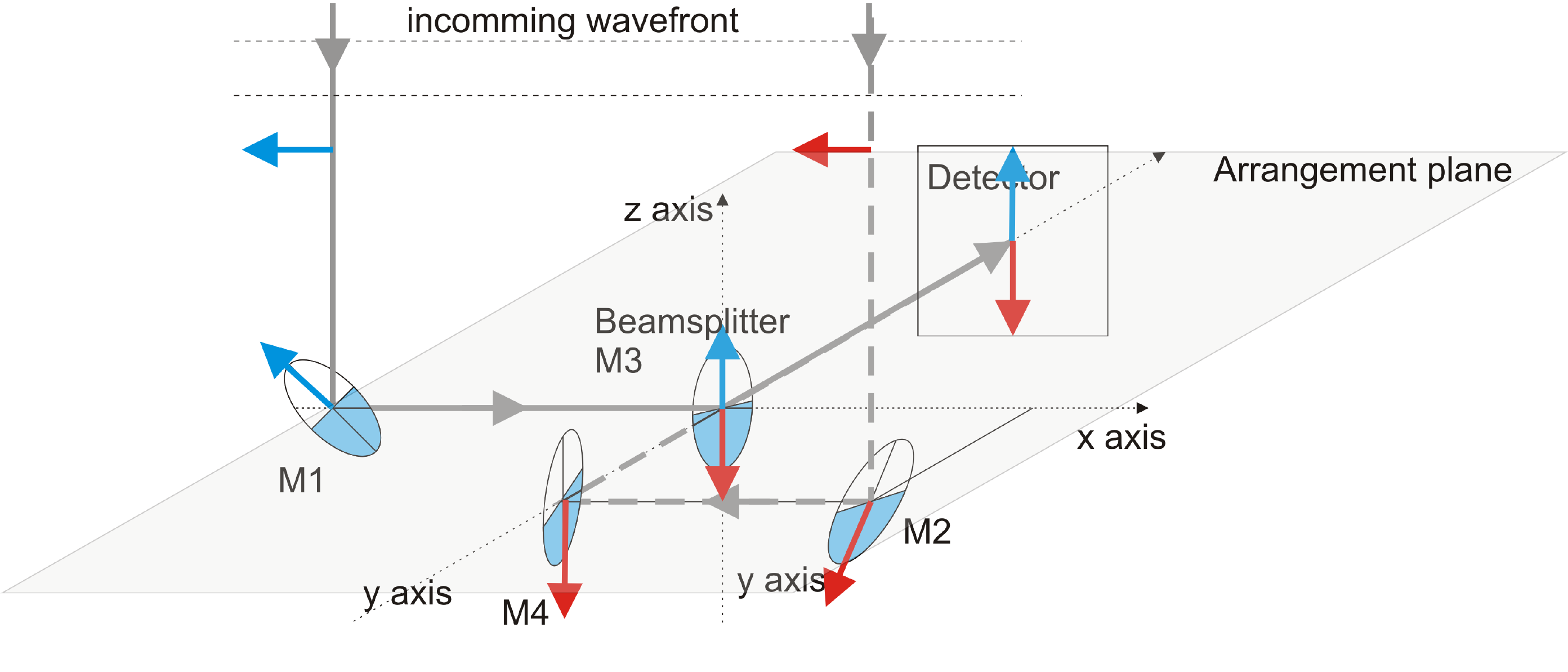}
\caption{\label{dt_1} The parallel transport of field directions 1
  along branch 1 and 2. It can be seen that one direction at the device
  entrance is parallel transported to two opposite directions whereby the direction depends on the path.}
\end{center}
\end{figure}

\begin{figure}
\begin{center}
\includegraphics[scale=0.32]{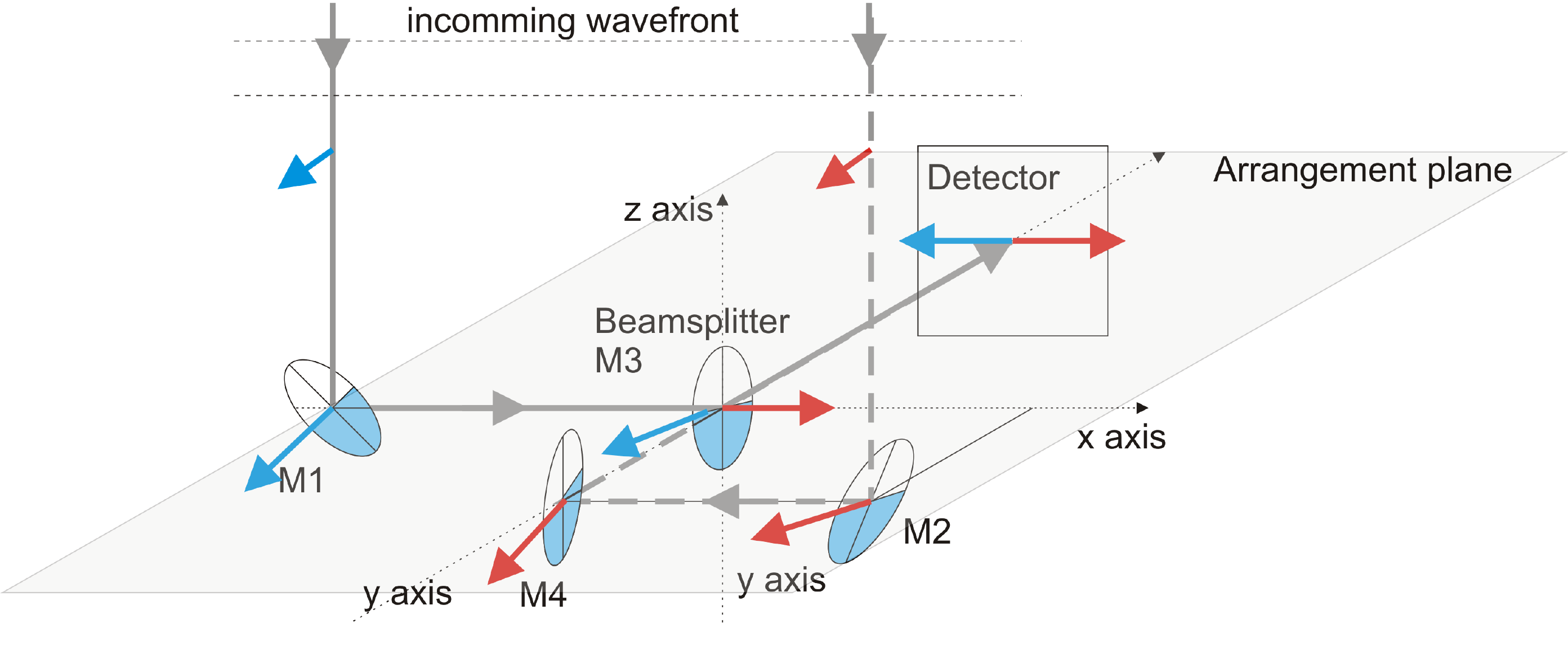}
\caption{\label{dt_2} The parallel transport of field directions 2
  along branch 1 and 2. It can be seen that one direction at the device
  entrance is parallel transported  to two opposite directions whereby the direction depends on the path.}
\end{center}
\end{figure}

Figure \ref{generic} shows the generic system for a 3D interferometer,
the simplest system being a wavefront division interferometer. For any
incoming plane wave this interferometer generates a superposition
between a plane wave propagated by branch 1 and 2, respectively. For
devices similar to the one shown in Figure \ref{generic} a central
beam can be found that has the property that the $\vec{k}$ vectors of
the outgoing beams originating from branch 1 and 2 are exactly
identical. Provided the path difference between branch 1 and 2 is
within the coherence length, this leads to either constructive or
destructive interference.

The beams in branch 1 and 2 have wave vectors called $\vec{k_1}$ and
$\vec{k_2}$, respectively. If the incoming beam is not a central beam
the propagated plane waves have a non vanishing wave vector component
in the detector plane, the latter being perpendicular to the central
beam.  The k-vector projections on this plane are denoted by capital
$\vec{K1}$ and $\vec{K2}$, respectively. The interferometer of Figure
\ref{generic} has the property

\begin{equation}
\label{eq:k_inverse}
\vec{K_1} = - \vec{K_2}
\end{equation}

We will further analyze the field transport in the generic
interferometer of Figure \ref{generic}. Two field
directions are consecutively attached to the central beam which are transported by
branch 1 and 2 (Figures \ref{dt_1}, \ref{dt_2}). The result by one
branch (1 or 2) is always contrary or opposite to the other one. This
is just the property of Equation (\ref{eq:k_inverse}) called 'k-flip'.

\begin{figure}
\begin{center}
\includegraphics[scale=0.32]{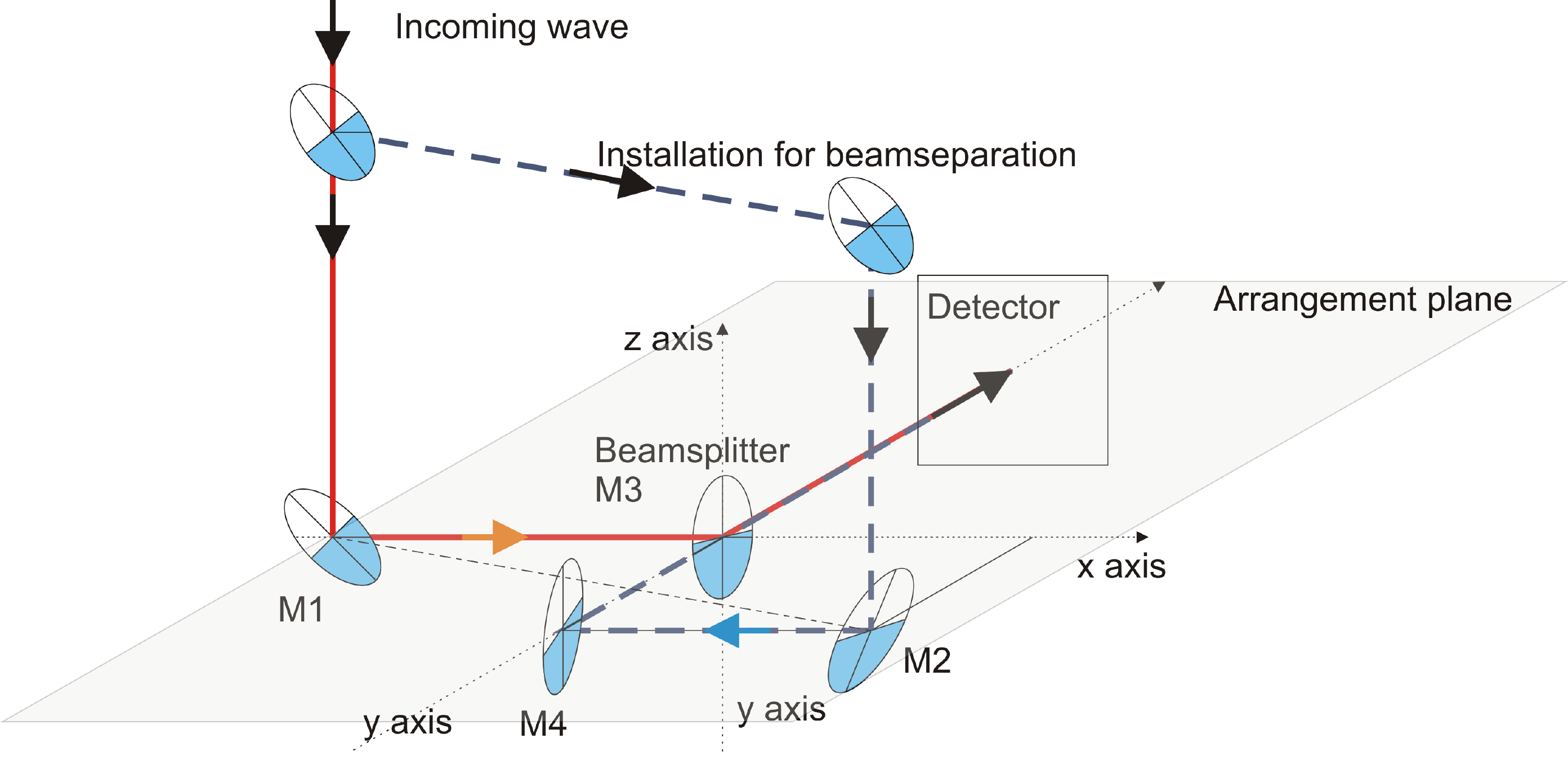}
\caption{\label{generic_ad} The generic 3D interferometer of Figure
  \ref{generic} combined with an installation for beam
  separation. This yields an interferometer with 'amplitude
  division'. The beam separation is achieved by two reflections
  displacing one of the two beams. It does not introduce a rotation in
  the field transport. Thus, the field rotation properties of Figure
  \ref{generic} are unchanged.}
\end{center}
\end{figure}

As can be expected from the introduction, this property is a result of
a 3D field transport. Wavefront division can be excluded as a cause as
shown by Figure \ref{generic_ad}.  The figure shows that
the incoming wave is split into two waves whereof one is displaced by
a shift. The two newly introduced mirrors are parallel and do not
introduce any rotation. As a result, the new set up is an
interferometer with amplitude splitting which has the 'k-flip'
property.

\begin{figure}
\begin{center}
\includegraphics[scale=0.32]{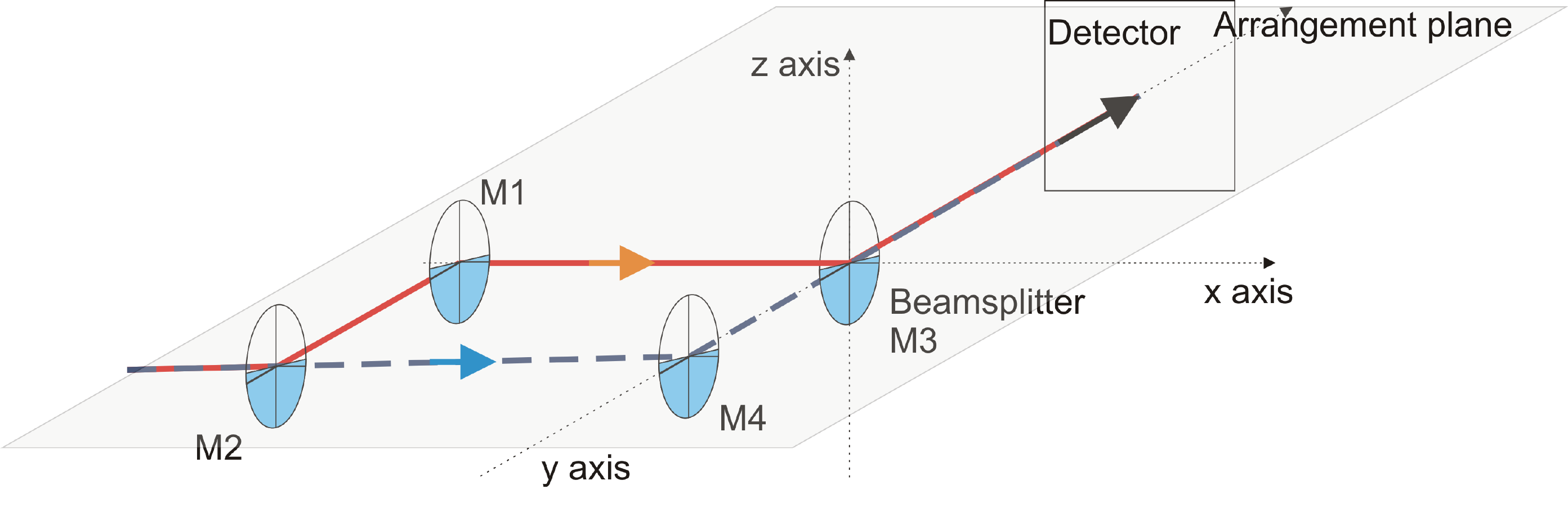}
\caption{\label{mz_4} Common Mach-Zehnder interferometer. The
  interferometer is a 2D interferometer.}
\end{center}
\end{figure}

\begin{figure}
\begin{center}
\includegraphics[scale=0.32]{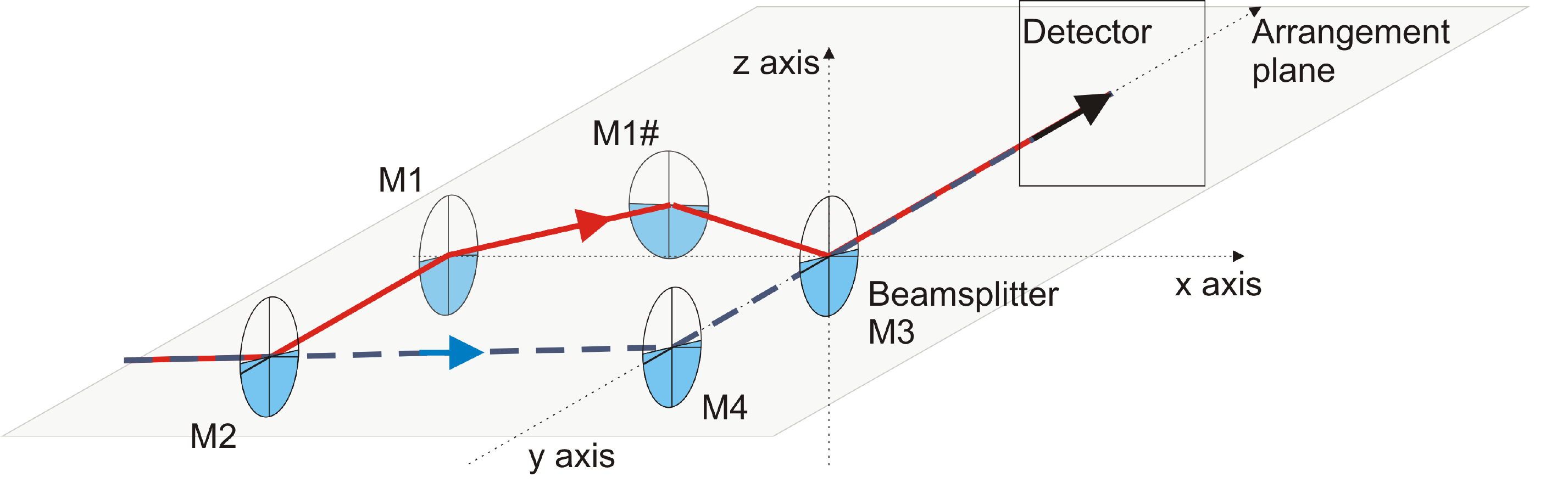}
\caption{\label{mz_5} The interferometer of Figure \ref{mz_4} with one
  additional mirror inserted into branch 2. The interferometer is a 2D
  interferometer.}
\end{center}
\end{figure}

\begin{figure}
\begin{center}
\includegraphics[scale=0.32]{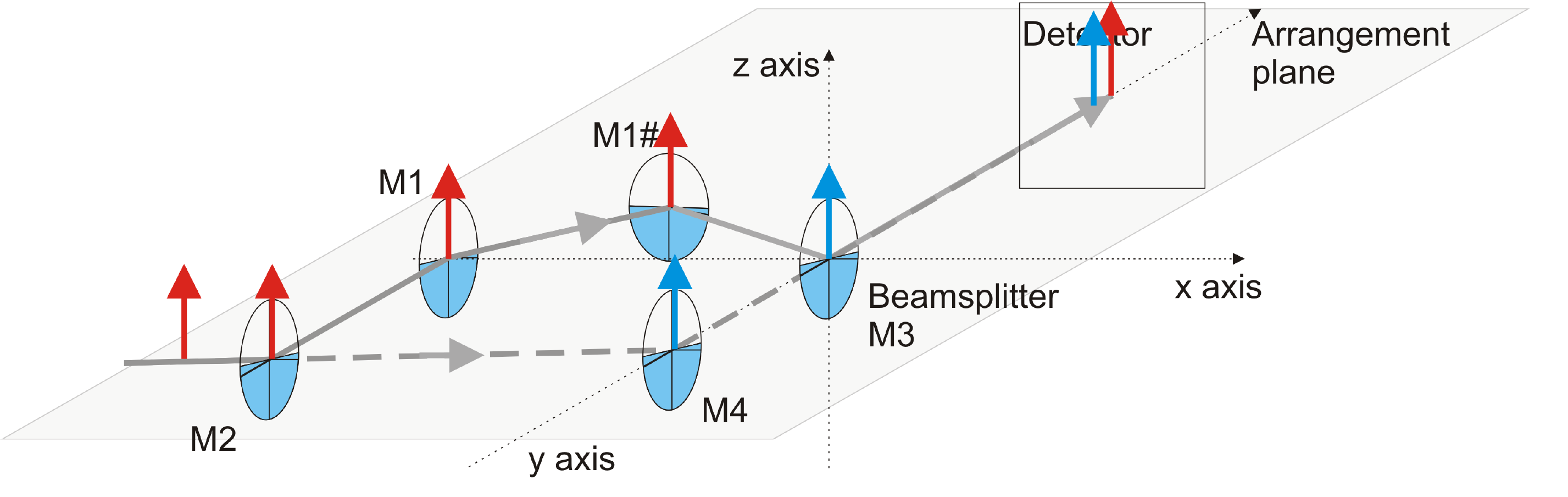}
\caption{\label{mz_5_ax} Parallel transport in the interferometer as
  shown in Figure \ref{mz_5}. The initial field direction is
  perpendicular to the arrangement plane. It can be seen that the
  parallel transport of the field direction does not lead to an
  inversion between the two branches.}
\end{center}
\end{figure}

\begin{figure}
\begin{center}
\includegraphics[scale=0.32]{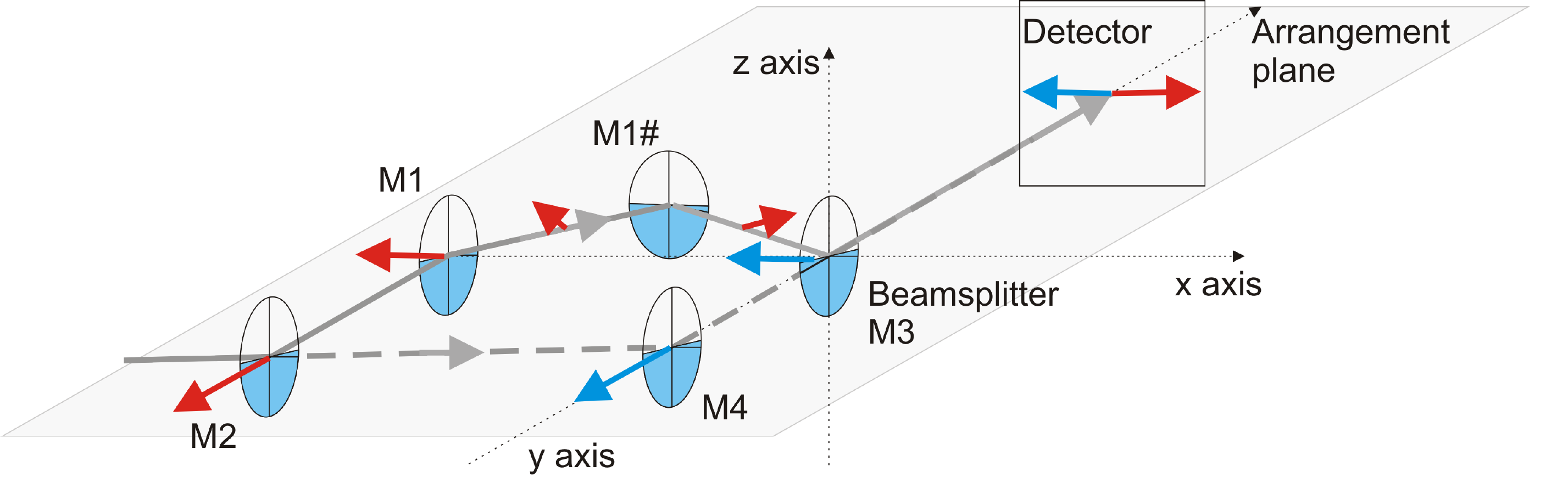}
\caption{\label{mz_5_bx} Parallel transport in the interferometer as
  shown in Figure \ref{mz_5}. The initial field direction is in the
  arrangement plane. It can be seen that the parallel transport of the
  field direction leads to an inversion between the two branches.}
\end{center}
\end{figure}

2D interferometers do not have this property. Figure \ref{mz_4} shows a
common 2D interferometer with two branches (a Mach-Zehnder
interferometer). Figure \ref{mz_5} shows the same type of
interferometer but with an additional mirror in one of the
branches. In fact, this is the only design freedom we have in 2
dimensions. 

It can be immediately seen that the Mach-Zehnder interferometer does
not have the 'k-flip' property. Insertion of an additional mirror
introduces an inversion of the field components in the arrangement
plane, indeed (Figure \ref{mz_5_bx}). Yet orthogonal to this plane the
situation remains unchanged (Figure \ref{mz_5_ax}). In conclusion,
Equation (\ref{eq:k_inverse}) does not hold for both directions.

\section{Evaluation for incoherent sources}
\label{incoh_source}

We study the interferogram for an object consisting of incoherent
emitters (sources) further. In addition, we assume that both branches
have the same transmission. Thus, both plane waves have the same
intensity. As we assume stationary, quasi-monochromatic
light~\cite{born}, each object point $s$ leads to two beams
which form an interference pattern with a real-valued, modulated intensity
$\mathfrak{I}$ given by

\begin{equation}
\label{eq:k_if}
\mathfrak{I}_s(\vec{x}) = a_s *  \cos \left( 2 \vec{K1_s} \cdot \vec{x} + \beta_s \right) 
\end{equation}

We have suppressed the constant term in Equation (\ref{eq:k_if}). 
$\vec{x}$ is a 2D coordinate vector in the detector plane, $\beta_s$
is a phase angle and $a_s$ is a real-valued intensity parameter. The index $s$
recalls the fact that the light comes from a source point called
'$s$'.

Assuming a set of mutually incoherent sources this yields~\cite{born}

\begin{equation}
\label{eq:inc_sum}
\mathfrak{I}(\vec{x}) = \sum_s I(s) \mathfrak{I}_s(\vec{x})
\end{equation}

$I(s)$ is the real-valued intensity of the light source $s$ as introduced by
\cite{born}. This expression has to be well distinguished from the van
Cittert-Zernicke theorem and the complex visibility $j(P_1,P_2)$ using
the notation of M. Born and E. Wolf~\cite{born}.  $P_1$ and $P_2$
denote the location of the two apertures in a 'division of wavefront'
setup. In Figure \ref{generic} this is the location of the mirrors M1
and M2 having the coordinates $P_1$ and $P_2$,
respectively. $j(P_1,P_2)$ is measured by testing the 2-point coherence property
at $P_1$ and $P_2$. In principle, this can be done by any
coherence sensitive experiment which compares light from $P_1$ and
$P_2$. It is the simplest approach to overlap the light coming from
$P_1$ and $P_2$ and to look for any constructive or destructive
interferences. Any visible interference pattern can be attributed to
the mutual coherence properties of the light from $P_1$ and
$P_2$. Concerning this, it is unimportant how this interference
pattern is actually produced. The Michelson stellar interferometer is
a famous example~\cite{born}. Alternatively, two different telescopes
can be used for $P_1$ and $P_2$, thereby increasing the light
sensitivity. Light wave variations over the experimentally finite
areas of $P_1$ and $P_2$ are filtered out, similar to a fundamental
mode filter. Although the finite extension of $P_1$ and $P_2$ is
undesirable for the evaluation, a finite extension is needed to collect
sufficient light. The light from $P_1$ and $P_2$ is brought to
interference at an angle of incidence on the detector($k \neq
0$). However, for all these cases, the k-vector of this interference
is imposed by the setup and is independent of $K1_s$. Furthermore, it
should be emphasized that $P_1$ has nothing to do with $\vec{x}$ in
Equation (\ref{eq:k_if}).

\begin{figure}
\begin{center}
\includegraphics[scale=0.32]{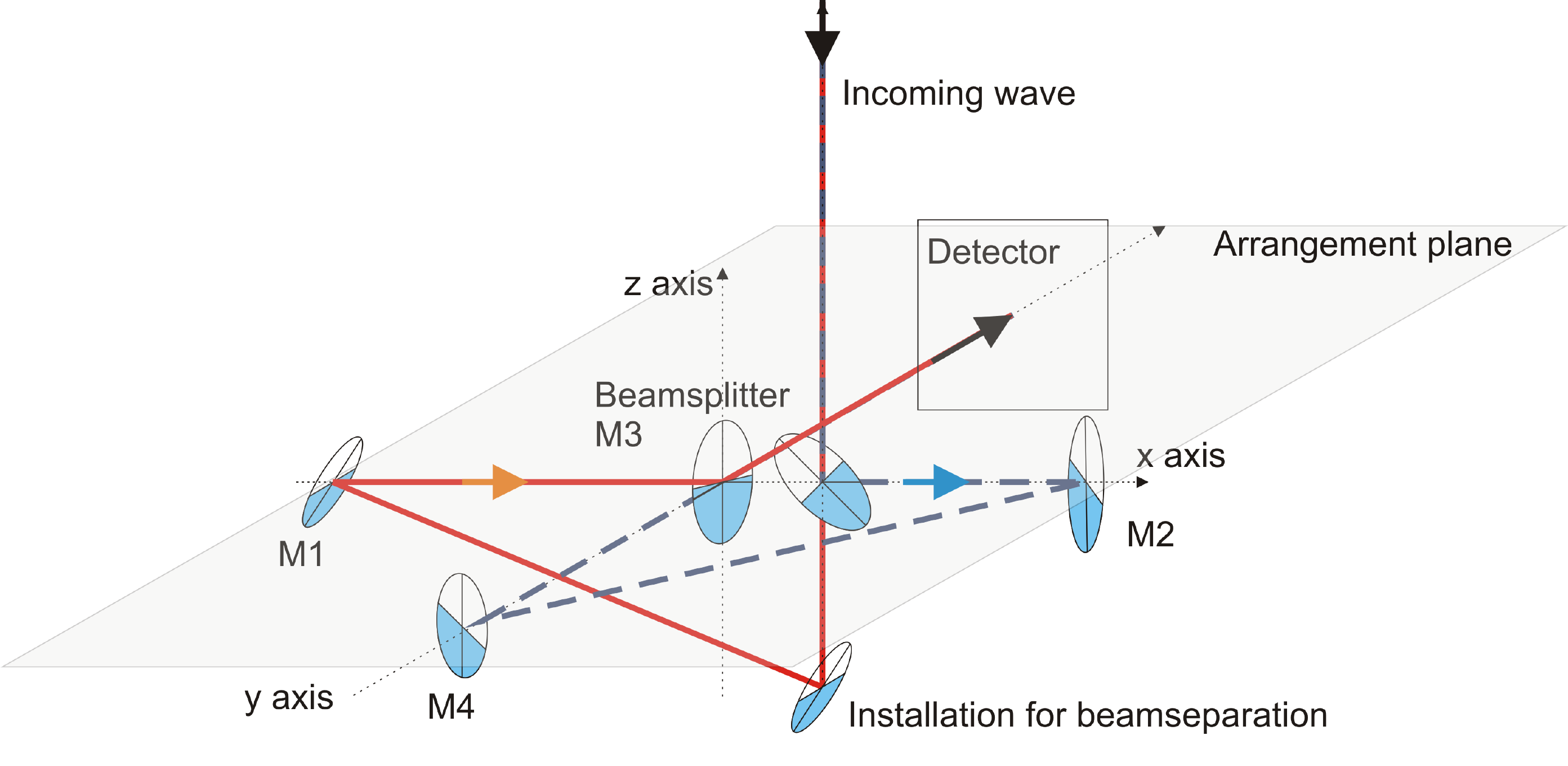}
\caption{\label{generic_ad2} The type of interferometer of Figure
  \ref{generic_ad} but with some rearrangement of the branches such as
  to give the setup a more 'elegant' appearance. Furthermore, the two
  branches are symmetric and of similar or equal length.}
\end{center}
\end{figure}

It is the purpose of the Michelson stellar interferometer and its
advanced multi-telescope successors to measure the size and location
of stars. The complex degree of coherence $j(P_1,P_2)$ must be
measured for many different pairs of $P_1$ and $P_2$ which together
yield the image. This can be done by actively changing the setup
(e.g. moving M1,M2) or by using either the earth or the satellite
rotation. This method is called aperture synthesis.

In conclusion, the angular resolution obtained by aperture synthesis
is not due to any particular 3D property of the interferometer but due
to different measurements with different orientations of the whole
interferometer in space.

The interference pattern recorded by an interferometer as shown in
Figure \ref{generic} is very different from the previously described
Michelson stellar interferometer since the former yields a real-valued
intensity pattern described by Equation (\ref{eq:inc_sum}) whereas the
latter produces a complex-valued 2-point correlation function
$j(P_1,P_2)$. The former depends on the modulation of the light field
over M1 and M2 while the latter does not measure this type of
property.

To exemplify this further, we assume equal transmission for all 's'
modes ($a_s = 1$). We call $I(s)$ the emission intensity of source
point $s$. Equation (\ref{eq:inc_sum}) can then be reformulated

\begin{equation}
\label{eq:inc_sum2}
\mathfrak{I}(\vec{x}) = \sum_s I(s) *  \cos \left( 2 \vec{K1_s} \cdot \vec{x} + \beta_s \right) 
\end{equation}

$K1_s$ and $\beta_s$ can be calculated, independently of the object,
for every light point $s$, provided the interferometer is
known. Alternatively the interferometer can be calibrated. If $\beta_s
= \beta$ for all s, i.e. if $\beta_s$ is independent of the index $s$,
the point $x=0$ is a point of stationary phase. The Equation
(\ref{eq:inc_sum2}) is a linear equation for $I(s)$. It is of course
interesting to invert this equation to get $I(s)$ from measured
results $\mathfrak{I}(\vec{x})$. The solution is unique under certain
conditions. If that is not the case, different measurement sets can be
evaluated together, the diffraction limit being a strict criterion for
uniqueness.

The evaluation has many similarities with Fourier transform
spectroscopy. The time or frequency variable in Fourier transform
spectroscopy corresponds to the space or $\vec{K1}$ vector
variable in our concept. An application of the Wiener-Khintchine theorem~\cite{born}
yields that the Fourier transform of the observed interference pattern
$\mathfrak{I}(\vec{x})$ corresponds to the real-valued spectral power
density $ I(\vec{K1_s}) := I(s) $. Here, we denote the light source
$s$ by its wave vector component $\vec{K1_s}$. This has some ambiguity
since it is not possible to distinguish $\vec{K1}$ from $ - \vec{K1}$
which can be seen from Equation (\ref{eq:k_if}) since $\vec{K1}$ and $
- \vec{K1}$ yield the same pattern. The problem does not exist in the
time domain, i.e. in classical Fourier transform spectroscopy as light
always has a positive frequency~\cite{born}. However, the object
in our setup usually consists of points to the 'right' and to the 'left' yielding
both signs of $\vec{K1}$. If this were not the case, the ambiguity
would not arise. In fact, this has some similarity with a carrier
phase method~\cite{yosh}. Alternatively, two data sets
$\mathfrak{I}(\vec{x})$ can be evaluated simultaneously with a slight
sub wavelength change of path length in one of the branches. The
fringe motion can be evaluated to distinguish between $\vec{K1}$ and $
- \vec{K1}$.

In conclusion, this shows that an appropriate 3D interferometer yields
valuable information about light emitting, incoherent objects. In the
coherent case, mutual interference between different points 's'
entails a different approach~\cite{berz} (Section \ref{coh_source}).

\begin{figure}
\begin{center}
\centerline{\includegraphics[scale=0.32]{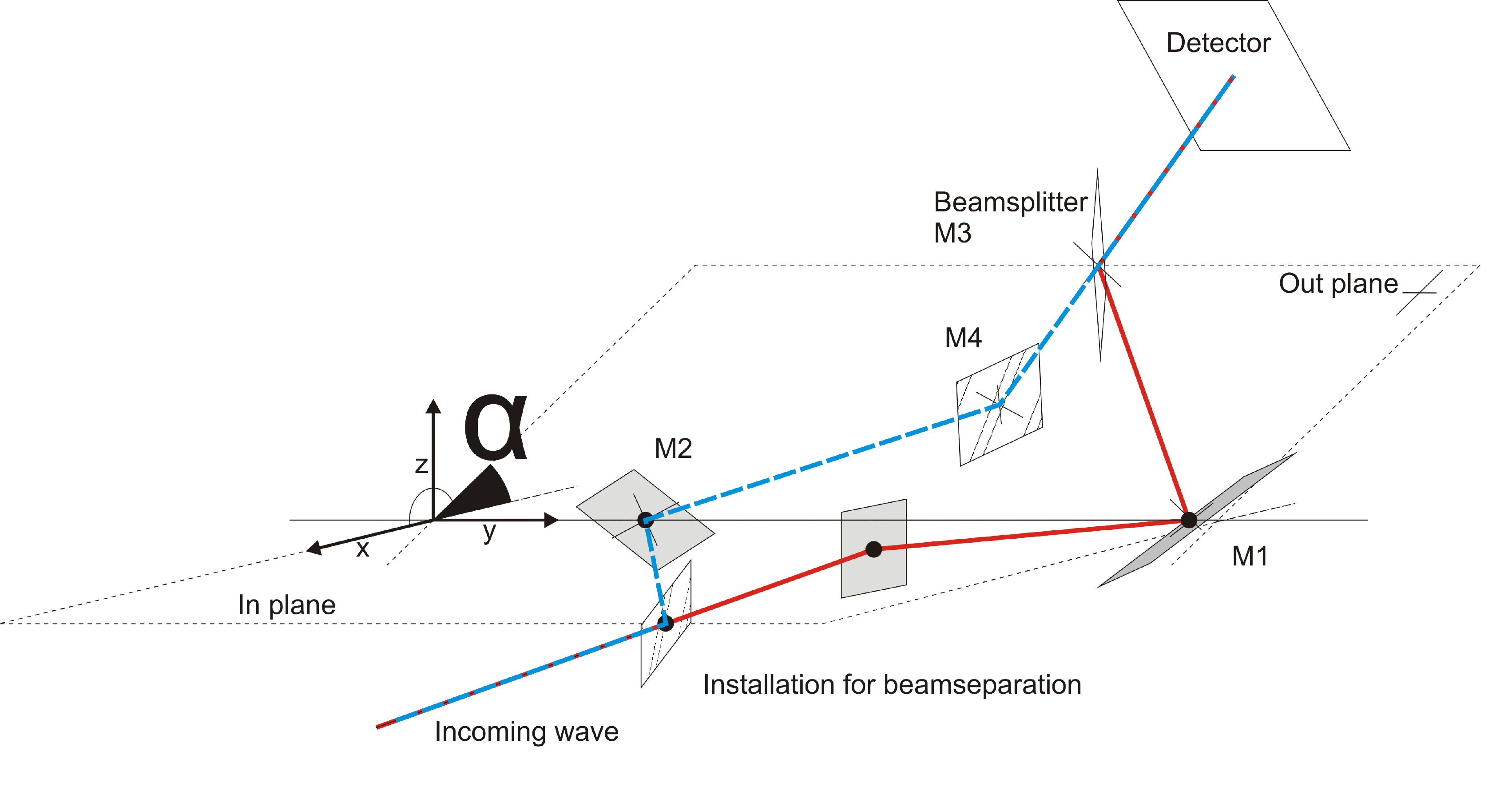}}
\caption{\label{gi_beta} A general field rotation. The twisting in 3
  dimensions leads to a mutual field rotation between the light fields
  propagated by branch 1 and 2.}
\end{center}
\end{figure}

\begin{figure}
\begin{center}
\centerline{\includegraphics[scale=0.32]{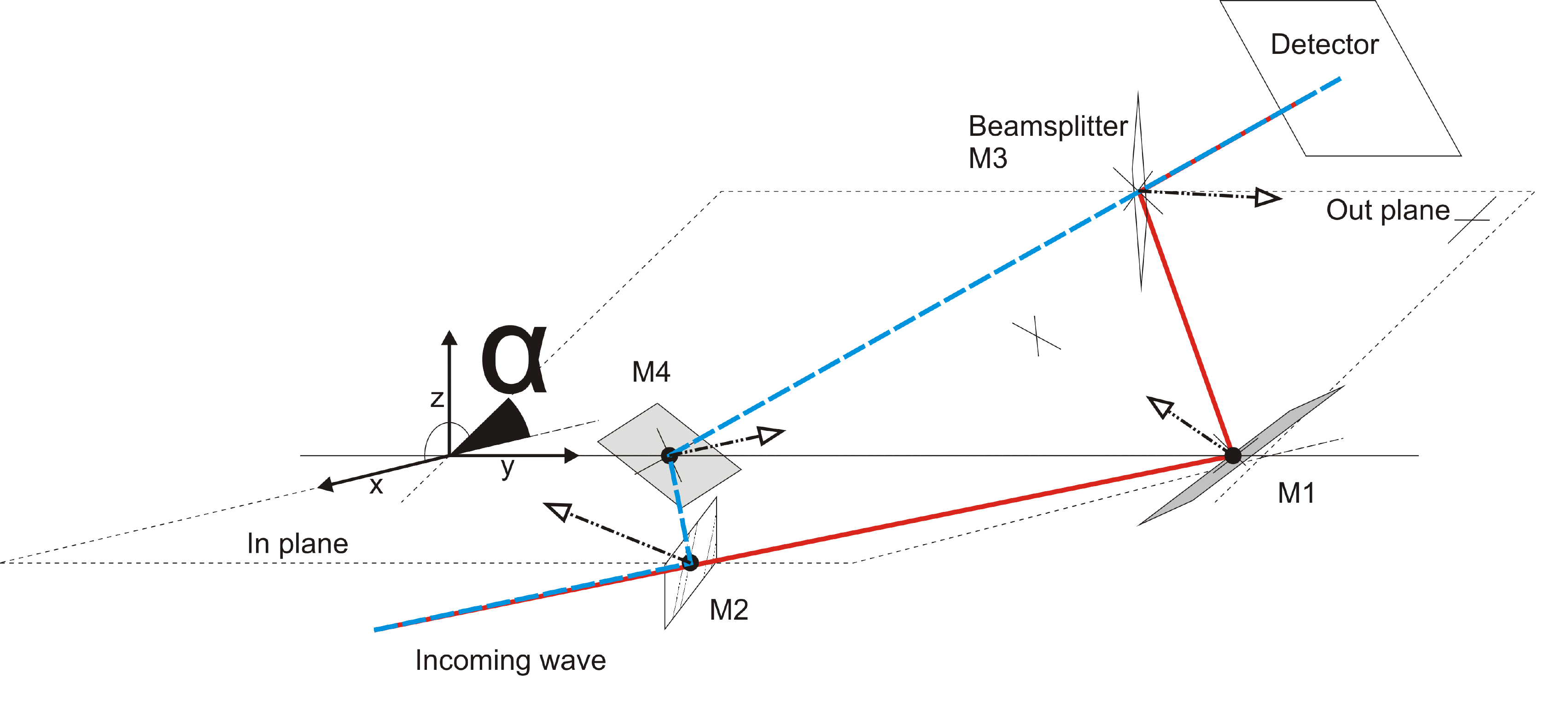}}
\caption{\label{3d_4mirror} A 3D interferometer with 4 beam
  reflections. In contrast to Figure \ref{gi_beta} such an
  interferometer does not lead to a mutual field rotation.}
\end{center}
\end{figure}

\section{Evaluation for coherent waves}
\label{coh_source}

In the case of spatially coherent light, the incident wave has a well
defined complex field $E(P)$ and the field values at different points
P are not only correlated but in a strict phase relation. In the case
of amplitude splitting (Figures \ref{generic_ad}, \ref{generic_ad2})
mutual phase relations  can be measured in the detector plane. As a
consequence, the complex field $E(P)$ can be determined.

In the most general case, two copies of the field called $E1$, $E2$
come to interference in the detector, $E1$ and $E2$ being the fields
propagated by branch 1 and 2, respectively. In the following, it will
be assumed that $E1$, $E2$ are the field functions defined in the
plane of the detector. For an appropriate interferometer, $E2$ is
determined by $E1$. The mapping can be determined by performing a back
propagation of field $E1$ by branch 1 to a plane before the
interferometer and  a subsequent forward propagation by branch 2 to get $E2$.
Thus, the mapping between $E1$ and $E2$ can include field propagation
and diffraction effects which can be calculated exactly by solving
Maxwell's equations (or e.g. the Helmholtz equation)~\cite{born}.
This yields

\begin{equation}
\label{eq:basic_U}
E2_{s} = \sum_{t} U_{s,t} E1_t =: U(E1) =: U E1 
\end{equation}

In this equation we introduce $s$ and $t$ as indices for different
points on the detector. U is a linear mapping between $E1$ and $E2$
which is a property of the interferometer and not of a particular
field.

The complex field $E1$ is obtained by solving a linear equation if
U is known and if the interference and the amplitude of $E1$ are known
from prior knowledge~\cite{berz}\cite{berz2}.

It is interesting to note that the incoherent case yields the
real-valued power density $ S(\vec{K1_s}) $ whereas the coherent case
yields the complex-valued field $E1$. These types of interferometers are
completely different.

\section{General 3D interferometer}
\label{gen_if}

\begin{figure}
\begin{center}
\includegraphics[scale=0.32]{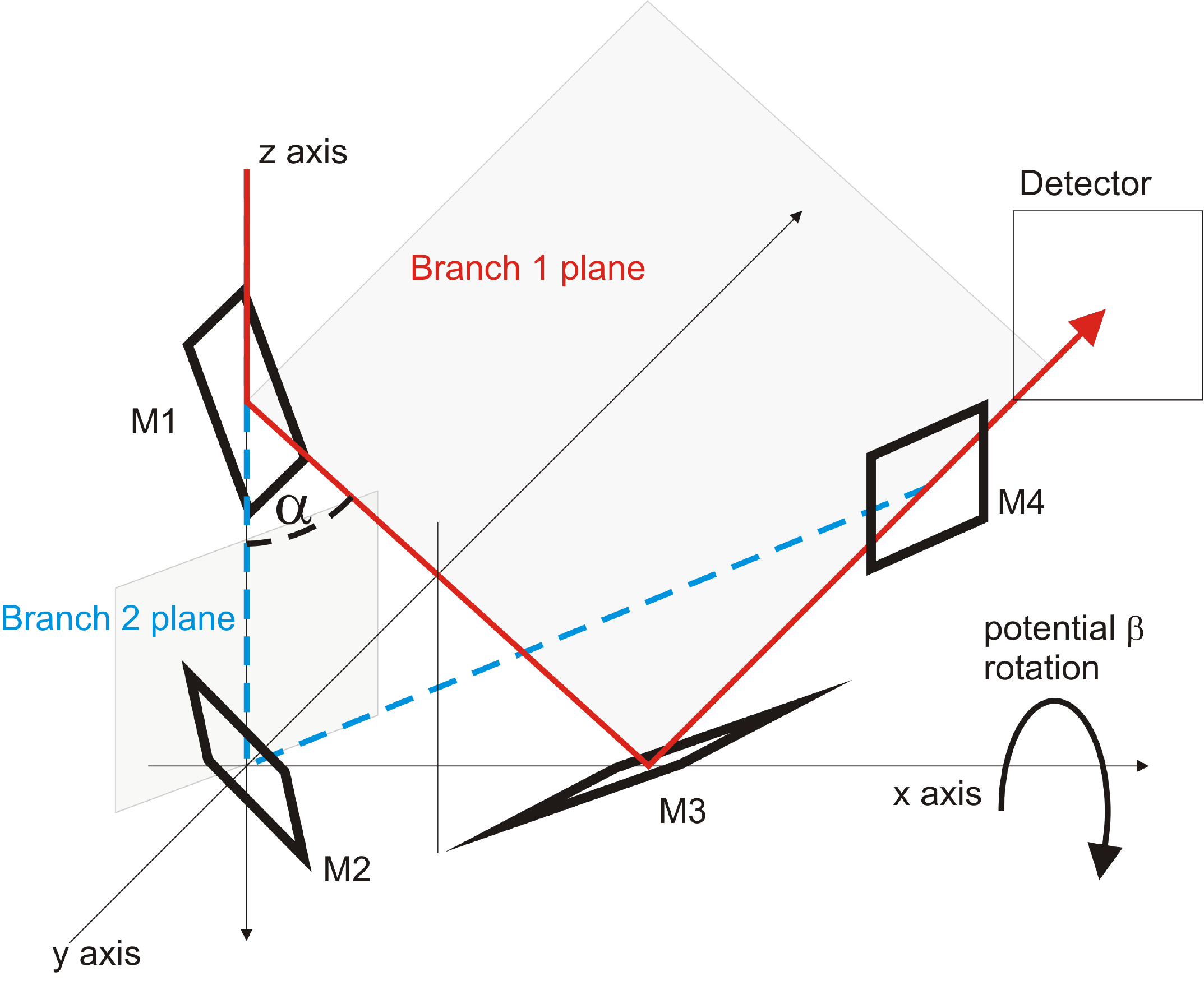}
\caption{\label{2m_1}The same type of interferometer as shown in
  Figure \ref{3d_4mirror} but shown for a $90^\circ$ folding of the
  interferometer planes. A plane 1 and 2 are depicted for branch 1 and
  2, respectively. This makes it easier to track the field
  directions. This is done in Figure \ref{2m_2} and \ref{2m_3}.}
\end{center}
\end{figure}

\begin{figure}
\begin{center}
\includegraphics[scale=0.32]{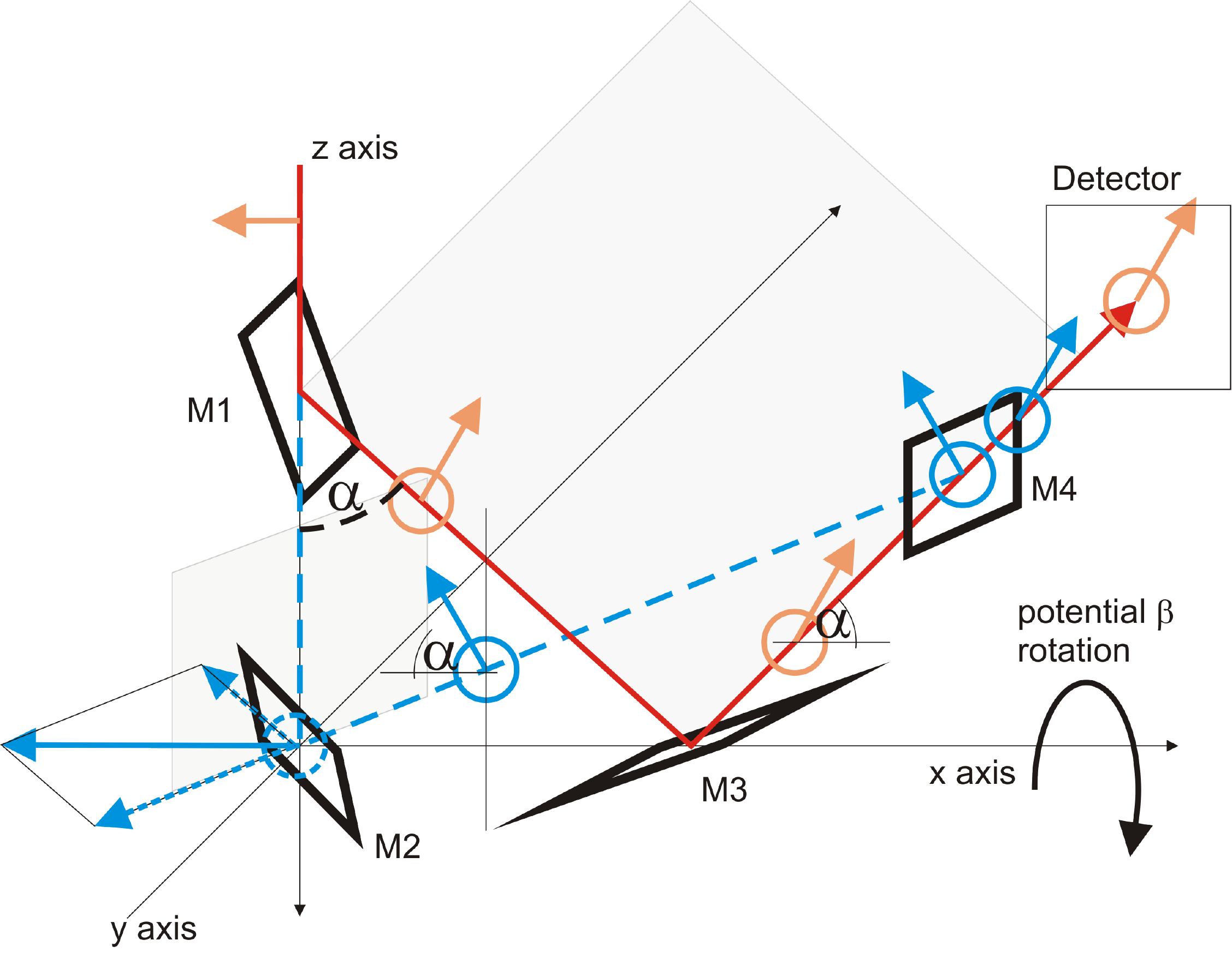}
\caption{\label{2m_2} The interferometer of Figure \ref{2m_1} with a
  depicted field direction. The parallel transport of this field
  direction is equal for branch l and 2. Thus, this type of equal path interferometer with 4 mirrors does not lead to a mutual field rotation.}
\end{center}
\end{figure}

\begin{figure}
\begin{center}
\includegraphics[scale=0.32]{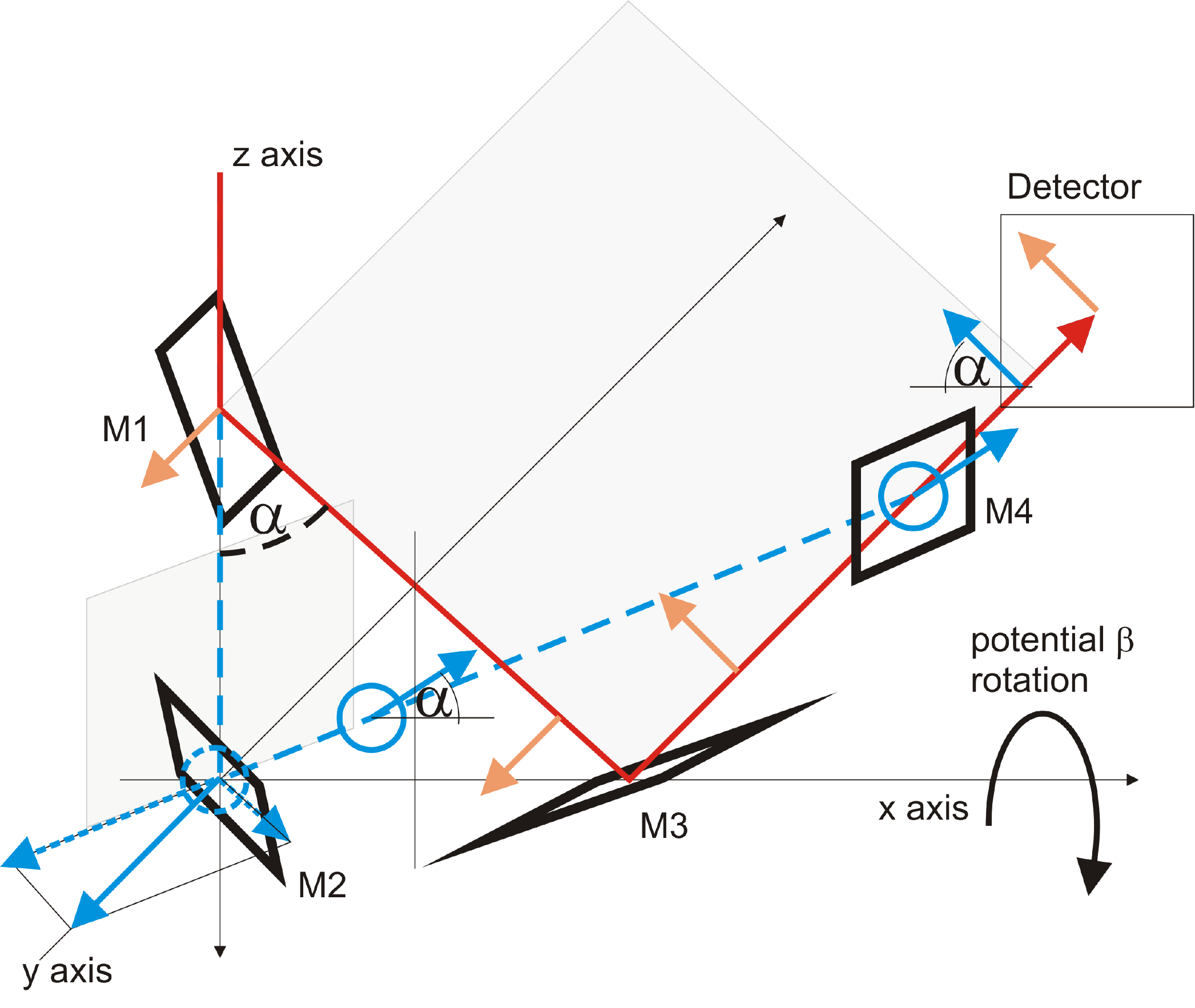}
\caption{\label{2m_3}The interferometer of Figure \ref{2m_1} with a
  depicted field direction orthogonal to Figure \ref{2m_2}. The
  parallel transport of this field direction is equal for both
  branches (l and 2). Thus, this type of equal path interferometer with
  4 mirrors does not lead to a mutual field rotation.}
\end{center}
\end{figure}

The mapping in Equation (\ref{eq:k_inverse}) corresponds to a
$180^\circ$ rotation called 'k-flip'. In the coherent case we
introduced a mapping U which is more general. In particular, U might
represent rotations other than $180^\circ$. Parallel transport can
generate such rotations as shown in Figure \ref{gi_beta}. The
interferometer consists of six wave reflections. Two reflections are
performed by beam splitters at the entrance and the exit of the
interferometer, four reflections are exerted by mirrors. The
interferometer is characterized by an in-plane and an out-plane. The
in-plane is defined by the transmitted and reflected beam on the input
beam splitter. Referring to the output beam splitter, an out-plane can
similarly be defined.  The two planes form an angle $\alpha$. Hence,
they are not identical. If $\alpha$ becomes $90^\circ$ the
interferometer in Figure \ref{gi_beta} is transformed into an
interferometer as displayed in Figure \ref{generic_ad}, being actually
a generalization of the latter.

For $\alpha = 0$ the field rotation in U becomes zero, for $\alpha =
90$ the field rotation in U is $180^\circ$. For all angles in between,
smaller field rotations are generated. The mapping U is a rotation which
entails the existence of a unique fixed point. This property is called
the 'U-property' and replaces the 'k-flip' property known
from the incoherent case in the coherent case.

It is interesting to note that the same cannot be achieved for an
equal path interferometer with just four reflections as shown in Figure
\ref{3d_4mirror}. The set of possible interferometer configurations is
described by an ellipsoid with the two beam splitters at the foci. We
analyze the four reflection setup in Figure \ref{2m_1} using an
interferometer which is folded by an angle of $\beta = 90^\circ$. Holding
$\beta$ fixed the angle $\alpha$ can be varied. Figures \ref{2m_2} and
\ref{2m_3} show that parallel transport does not lead to a mutual
field rotation. It can be anticipated that this remains true for
$\beta < 90^\circ$. For $\beta = 0^\circ$ we have a flat Mach-Zehnder
interferometer. Of course such an interferometer does not possess a
field rotation. In addition to that, we have verified by a numerical
simulation that the field rotation is really zero for all intermediate
$\beta$ values. This actually proves the initial statement that a four
reflection, equal path interferometer does not possess a field
rotation.

This might be condensed into the statement that non-trivial, parallel
transport only exists for a twisted interferometer with more than
four mirrors whereas a folded interferometer with four reflections does not possess
a mutual field rotation.

\section{Concluding remarks}

We have shown that an appropriate 3D interferometer possesses a mutual
field rotation as observed in an interference between the field of branch
1 and 2. Thereby, a 3D interferometer is defined by one of the
following definitions:

\begin{itemize}
  
\item{An interferometer has the 3D property if the central beams do not
  lie in a plane, not even approximately (Type I).}

\item{An amplitude division interferometer has the 3D property if the
entrance plane given by the central beam at the first beamsplitter and
the outgoing plane given by the central beams at the second
beamsplitter are different (Type II).}
\end{itemize}

The two definitions of 3D are essentially equal but the Type II
property cannot be used for a wavefront division interferometer such
as displayed in Figure \ref{generic}.

The rotation properties can be expressed as a property of plane waves
which are propagated differently by branch 1 and 2. The projection of
the wave vectors on the detector plane denoted $\vec{K_1}$ and
$\vec{K_2}$ have the 'k-flip' property

\begin{equation}
\label{eq:k_inverse_concl}
\vec{K_1} = - \vec{K_2}
\end{equation}

For coherent light, the complex light fields are defined, $E1$ and
$E2$ being the fields from branch 1 and 2, respectively. Therefore, a
mapping U can be determined which describes the transformation in the
interferometer, in particular the mapping from field $E1$ to field $E2$.

\begin{equation}
\label{eq:k_inverse_concl2}
E2 = U ( E1 )
\end{equation}

If U has just one fixed point, the interferometer is said to have the
'U-property'.

If the light sources consist of incoherent emitters no field $E1$ is
defined and consequently no light field phase can be determined. The evaluation is
based on an application of the Wiener-Khintchine theorem~\cite{born}
which yields the real-valued power spectrum $ S(\vec{K1_s})$ of the
$\vec{K1_s}$ distribution in the radiation field.  Hence, an
interferometer with the 'k-flip' property is needed.

On the other hand, coherent fields can be analyzed by an interferometer
with the 'U-property'.  The evaluation of the generated interferograms
has become feasible due to the recent discovery of linear phase
retrieval in SRI interferometers~\cite{berz}.

As a consequence, depending on the radiation field (coherent, incoherent)
a different physical interferometer and a different evaluation method
has to be used.

For this, the two properties ('k-flip' or 'U-property') are very
useful and can be generated by an appropriately designed, twisted
interferometer using parallel field transport.


%

\end{document}